\documentclass[11pt, onecolumn]{article}

\usepackage{fancyhdr}
\usepackage[english]{babel}
\usepackage{abstract}

\usepackage{amsmath}  
\usepackage{amssymb} 
\usepackage{mathrsfs}

\usepackage{euscript}
\usepackage{hyperref}

\usepackage{cyr}
\usepackage{epsfig}
\usepackage{epstopdf}

\usepackage{titlesec}

\makeatletter
\renewcommand{\@oddhead}{}
\renewcommand{\@oddfoot}{\hfill ---~\thepage~---\hfill}
\makeatother

\titleformat*{\section}{\Large\bfseries}
\titleformat*{\subsection}{\large\bfseries}

\fancypagestyle{firststyle} 
{
\fancyhead[L]{\small Published in Astronomy Letters, 2014, Vol. 40, No.7, pp. 425-434 \\ original russian text: Pis'ma v Astronomicheskii Zhurnal, 2014, Vol. 40, No. 7, pp. 473-483.\hfill}
\fancyfoot[L]{\hfill ---~\thepage~---\hfill}
 
}

\begin{document}
\thispagestyle{firststyle}

\begin{center}
\Large H$_{2}$O Maser Pumping: The Effect of Quasi-Resonance Energy Transfer in Collisions between H$_2$ and H$_2$O Molecules

\vspace{0.5cm}
\large A.V. Nesterenok $^{1 *}$ and D.A. Varshalovich $^{1,2}$
\vspace{0.5cm}

\normalsize $^1$ Ioffe Physical-Technical Institute, Politekhnicheskaya St. 26, Saint~Petersburg, 194021 Russia

$^2$ St. Petersburg State Polytechnical University, Politekhnicheskaya St. 29, Saint~Petersburg, 195251 Russia

$^*$ e-mail: alex-n10@yandex.ru
\end{center}

\begin{abstract} 
\noindent
The effect of quasi-resonance energy transfer in collisions between H$_2$ and H$_2$O molecules in H$_2$O maser sources is investigated. New data on the state-to-state rate coefficients for collisional transitions for H$_2$O and H$_2$ molecules are used in the calculations. The results of ortho-H$_2$O level population inversion calculations for the 22.2-, 380-, 439-, and 621-GHz transitions are presented. The ortho-H$_2$O level population inversion is shown to depend significantly on the population distribution of the para-H$_2$ $J = 0$ and $2$ rotational levels. The possibility of quasi-resonance energy transfer in collisions between H$_2$ molecules at highly excited rotational-vibrational levels and H$_2$O molecules is considered. The quasi-resonance energy transfer effect can play a significant role in pumping H$_2$O masers in the central regions of active galactic nuclei and in star-forming regions.
\end{abstract}

Keywords: \textit{cosmic masers, star-forming regions, active galactic nuclei.}
\smallskip

DOI: 10.1134/S1063773714070068

\section*{Introduction}
H$_2$O maser emission is observed in many astrophysical objects: in the expanding envelopes of late-type stars, star-forming regions, and the central regions of active galactic nuclei (AGNs). The isotropic luminosities in the 22.2-GHz line for sources observed in our Galaxy lie between $\lesssim 10^{-7}$
and $10^{-2} L_{\odot}$, reaching 1$L_{\odot}$ in rare cases (W49N) (Palagi et al. 1993; Liljestr$\ddot{o}$m et al. 1989). The brightest Galactic H$_2$O masers are observed in star-forming regions. H$_2$O maser emission in star-forming regions is generated in circumstellar disks and the shocks produced by bipolar outflows and stellar winds from young stars (Torrelles et al. 2005). 

The first 22.2-GHz H$_2$O maser sources discovered in other galaxies had isotropic luminosities of 0.1--1 $L_{\odot}$ (Churchwell et al. 1977). In 1979, however, dos Santos and L$\acute{e}$pine (1979) detected intense H$_2$O maser emission from the central region of the galaxy NGC 4945. The isotropic luminosity of the maser was $\sim$ 100 $L_{\odot}$, which is higher than the typical luminosities of H$_2$O maser sources in our Galaxy by several orders of magnitude. About 150 galaxies with intense H$_2$O maser emission observed from their central regions are known at present (Tarchi 2012). The isotropic luminosities of the sources in specific cases exceed $10^4 L_{\odot}$ (Castangia et al. 2011). All these galaxies have evidence of activity in the galactic nuclei. H$_2$O maser emission in the central regions of AGNs is generated in accretion disks around supermassive black holes (Moran 2008). In specific cases,
H$_2$O maser emission can be observed along the radio jet (Peck et al. 2003) and can be associated with the outflows of dense molecular gas from the central engine (Greenhill et al. 2003). H$_2$O maser emission is a unique tool for investigating the structure and kinematics of the gas in the neighbourhoods of the central engines in AGNs.

As a rule, the collisional excitation of H$_2$O molecules to higher lying levels followed by the radiative de-excitation of these levels is considered as the main H$_2$O-maser pumping mechanism (Strelnitskii 1973; Yates et al. 1997). Varshalovich et al. (1983) proposed an H$_2$O-maser pumping mechanism in which the H$_2$O levels are excited through quasi-resonance energy transfer in collisions between H$_2$ and H$_2$O molecules. The absorption of ultraviolet radiation, the collisions of molecules with high-energy photoelectrons, and formation reactions lead to the population of highly excited rotational-vibrational molecular levels. The listed processes can take place in photo- and X-ray dissociation regions as well as in dissociative shocks. The de-excitation of excited molecular levels can occur both through the emission of electromagnetic radiation and through collisional processes. The quasi-resonance excitation energy transfer in collisions between molecules is possible.

In this paper, we investigate the effect of quasi-resonance energy transfer in collisions between H$_2$ and H$_2$O molecules on the H$_2$O level population inversion. The collisional rate coefficients from Dubernet et al. (2009) and Daniel et al. (2010, 2011) were used. In our calculations, we took into account the state-to-state collisional transitions for the lower 45 ortho-H$_2$O rotational levels and the para-H$_2$ $J = 0$ and 2 levels. The simple model of a gas-dust cloud with constant physical parameters was considered. In our calculations of level populations, we used the accelerated $\Lambda$-iteration method (Rybicki and Hummer 1991).

\section*{Quasi-resonance Energy Transfer in Collisions between H$_2$O and H$_2$ Molecules}

The main quasi-resonance conditions in collisions between molecules are: the Massey criterion, the conservation of the particle total rotational angular momentum, the conservation of the product of the parities of the particle wave functions, and the nuclear spin conservation for each of the molecule (Varshalovich et al. 1983).

According to the Massey criterion, the increment in the sum of the excitation energies of the molecules
before and after their collision must satisfy the inequality $\vert \Delta \varepsilon \vert \lesssim \hbar v/\rho$, where $\vert \Delta \varepsilon \vert$ is the excitation energy defect, $v$ is the relative velocity of the particles, and $\rho$ is their characteristic interaction radius. For
collisions between H$_2$O and H$_2$, this condition can be rewritten as

\begin{equation}
\vert \Delta \varepsilon \vert \lesssim \frac{\hbar v_T}{\rho} \simeq 4 \times 10^{-16}\,\sqrt{T_g} \;\; \text{erg},
\label{eq1}
\end{equation}

\noindent
where $v_T$ is the mean relative velocity of the molecules, $T_g$ is the kinetic temperature of the gas in kelvins; in our estimation, the H$_2$O-H$_2$ interaction radius $\rho$ was assumed to be $2.5 \text{\AA}$ (Valiron et al. 2008). The excitation energy defect is

\begin{equation}
\Delta \varepsilon = \varepsilon_{H_2O} + \varepsilon_{H_2} - \varepsilon'_{H_2O} - \varepsilon'_{H_2}, 
\nonumber
\end{equation}

\noindent
where $\varepsilon$ and $\varepsilon'$ are the level energies of the molecules before and after their collision, respectively.

The conditions for the conservation of the total angular momentum of the molecules and the conservation
of the product of the parities can be written as (Varshalovich et al. 1983)

\begin{equation}
\textbf{J}_{H_2} + \textbf{J}_{H_2O} = \textbf{J}_{H_2}' + \textbf{J}_{H_2O}', \quad \pi_{H_2} \pi_{H_2O} = \pi_{H_2}' \pi_{H_2O}',
\label{eq2}
\end{equation}

\noindent
where $\textbf{J}$ and $\textbf{J}'$ are the rotational angular momentum vectors of the molecule before and after the collision, $\pi$ and $\pi'$ are the parities of the molecule's wave function with respect to spatial coordinate inversion before and after the collision, respectively (the change of orbital angular momentum of the relative motion of the molecules in this case is 0). The parity of the wave functions is $\pi_{H_2} = (-1)^J$ for the H$_2$ molecule and $\pi_{H_2O} = (-1)^{J+K_a+K_c}$ for the H$_2$O molecule, where $K_a$ and $K_c$ are the quantum numbers that characterize the projections of the rotational angular momentum vector onto the molecule's internal axes (Tennyson et al. 2001). The nuclear spin conservation for the molecules means that the transitions changing the ortho/para states of the molecules are forbidden:

\begin{equation}
I_{H_2O} = I_{H_2O}', \quad I_{H_2} = I_{H_2}',
\label{eq3}
\end{equation}

\noindent
where $I$ and $I'$ are the nuclear spins of the molecules before and after their collision, $I = 0$ and 1 correspond to the para- and ortho-states, respectively.

Dubernet et al. (2009) provide the rate coefficients for transitions in collisions between ortho-H$_2$O and para-H$_2$ molecules for the lower 45 H$_2$O rotational levels and for H$_2$ in the $J = 0$ state with the change in rotational angular momentum $\Delta J = 0, +2$ as well as for H$_2$ in the $J = 2$ state with $\Delta J = 0, -2$. Consider the collisional H$_2$O transitions from level $i$ to level $k$ and the H$_2$ transitions from the $J = 2$ level to the $J = 0$ level, for which the quasi-resonance conditions (1)-(3) are met. We will denote the corresponding collisional rate coefficients by  $C_{ik}^{J = 2 \to 0}$, with $i < k$. Figure 1 presents the ratios of the collisional rate coefficients for quasi-resonance H$_2$O excitation and their mean values for the transitions without any quasi-resonance, i.e., $C_{ik}^{J = 2 \to 0}/ \langle C_{ik}^{lm} \rangle$. The averaging in $\langle C_{ik}^{lm} \rangle$ is performed over the following $l \to m$ transitions in H$_2$: $J = 0 \to 0$, $0 \to 2$, $2 \to 2$. The mean ratios of the rate coefficients are approximately 27, 13, 3, and 2 for gas temperatures of 300, 400, 800, and 1500 K, respectively.

\begin{figure}[t]
\centering
\includegraphics[width = 0.75\textwidth]{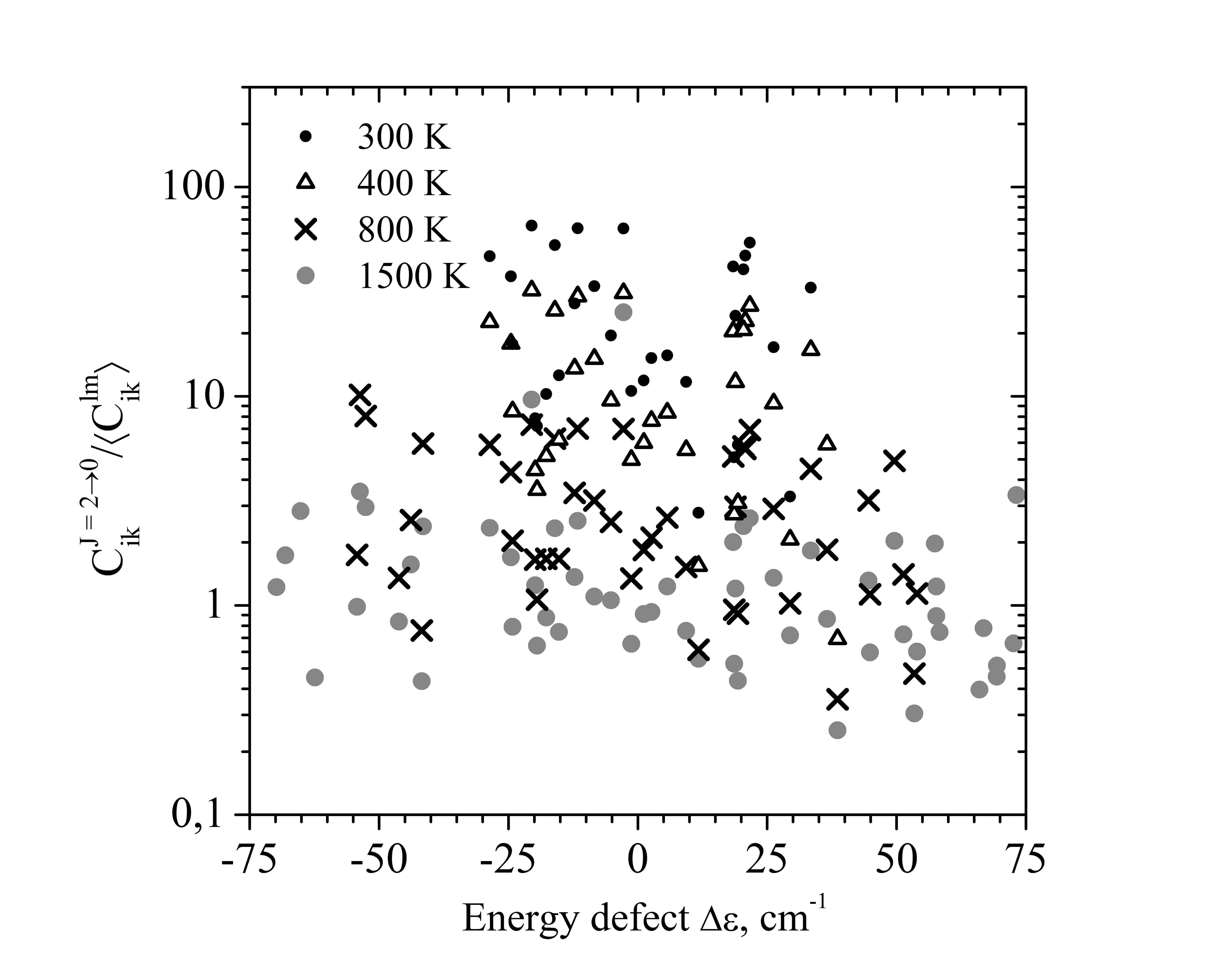}
\caption{\small{Ratios of the collisional rate coefficients for the H$_2$O and H$_2$ transitions for which quasi-resonance energy transfer takes place and their mean values for the transitions without any quasi-resonance. The excitation energy defect is along the horizontal axis. The data are presented for four temperatures (300, 400, 800, and 1500 K).}}
\end{figure}

\section*{Model Parameters}

The observed sizes of H$_2$O maser sources in star-forming regions are $\sim 10^{12}-10^{13}$ cm (Matveyenko et al. 2000; Uscanga et al. 2005; Matveyenko and Demichev 2010). To all appearances, the sizes of maser sources in accretion disks in the central regions of AGNs are $\sim 10^{14}$ cm (Kartje et al. 1999). The gas densities in H$_2$O maser sources can lie within the range $10^7 - 10^{10}$ cm$^{-3}$. At higher densities, there is thermalization of the level populations due to collisional processes. To all appearances, lower gas densities are also unlikely, because a high relative H$_2$O abundance is needed in this case. High H$_2$O column densities along the line of sight, $10^{18}-10^{19}$ cm$^{-2}$, are needed for the generation of intense H$_2$O maser emission (Yates et al. 1997). The gas temperature in maser sources is 300-1500 K. At higher temperatures, there is dissociation of H$_2$O molecules (Nesterenok and Varshalovich 2011). The collisional maser pumping is most efficient when the dust temperature is much lower than the gas temperature (Bolgova et al. 1977; Yates et al. 1997; Babkovskaia and Poutanen 2004). 

Here, we consider a one-dimensional plane parallel gas-dust cloud structure. The cloud sizes along two coordinate axes are much larger than that along the third $z$ coordinate axis. As a rule, the model of a flat cloud is considered for modeling the H$_2$O maser emission in the envelopes of late-type stars and in star-forming regions (Nesterenok et al. 2013; Hollenbach et al. 2013). 

The physical parameters adopted in our calculations are given in the table.

~\\
~\\
\begin{tabular}{l@{\quad\quad}|@{\quad\quad}l}
\multicolumn{2}{l}{\large\bf Table} \\ [5pt]
\hline \\ [-2ex]
Cloud half-thickness & $H = 10^{13}$ cm \\ [5pt]
Total number density of hydrogen atoms & $10^7$ cm$^{-3} \leq N_\text{H,\,tot} \leq 10^{10}$ cm$^{-3}$ \\ [5pt]
Ratio of H$_2$ and H number densities & $N_{\text{H}_2}/ N_\text{H} = 1$ \\ [5pt]
Relative He abundance & $N_\text{He}/N_\text{H,\,tot} = 0.1$ \\ [5pt]
Number density of ortho-H$_2$O molecules & $N_{\text{H}_2\text{O}} = 5 \times 10^{4}$ cm$^{-3}$ \\ [5pt]
Ortho-H$_2$O column density (perpendicular to cloud plane) & $\mathscr{N}_{\text{H}_2\text{O}} = 2H N_{\text{H}_2\text{O}} = 10^{18}$ cm$^{-2}$ \\ [5pt]
Gas temperature & 300 K $\leq T_g \leq 1500$ K \\ [5pt]
Dust temperature & $T_d = 100$ K \\ [5pt]
Turbulent velocity & $v_{\text{turb}}$ = 1 km s$^{-1}$ \\[5pt] \hline
\end{tabular}
~\\
~\\
The total number density of hydrogen atoms is $N_\text{H,\,tot} = N_\text{H} + 2N_{\text{H}_2}$
~\\

\section*{Calculating the H$_2$O Level Populations in a Gas-Dust Cloud}
\subsection*{Radiative Transfer in Molecular Lines}

We consider a gas-dust cloud that consists of a mixture of H$_2$ and H$_2$O molecules, H and He atoms, and dust particles. The physical parameters (the number densities of atoms and molecules, the dust content, the gas and dust temperatures) are assumed to be independent of the coordinates. However, the H$_2$O level populations are considered as functions of the $z$ coordinate.

In the one-dimensional geometry, the intensity of radiation $I$ at frequency $\nu$ depends on depth $z$ and angle $\theta$ between the $z$ axis and the radiation direction. The quantity $\mu = cos \theta$ is used instead of the variable angle $\theta$. The radiative transfer equation can be written as

\begin{equation}
\mu \frac{dI(z,\mu,\nu)}{dz} = -\kappa(z,\nu) I(z,\mu,\nu) + \varepsilon(z,\nu),
\label{eq4}
\end{equation}

\noindent
where $I(z,\mu,\nu)$ is the intensity of radiation at frequency $\nu$ in direction $\mu$, $\varepsilon(z,\nu)$ is the emission coefficient, and $\kappa(z,\nu)$ is the absorption coefficient. The
point $z = 0$ corresponds to the cloud boundary and the $z$ axis is directed into the cloud. The boundary condition for Eq. (\ref{eq4}) at $z = 0$ is $I(0,\mu,\nu) = 0$, where $\mu > 0$ (the directions into the cloud). At $z = H$, the boundary condition is taken to be $dI(z = H,\mu,\nu)/dz = 0$, corresponding to a cloud symmetric in physical parameters relative to $z = H$. Thus, the total cloud thickness is $2H$.

Each of the coefficients $\varepsilon(z,\nu)$ and $\kappa(z,\nu)$ is the sum of the emission or absorption coefficient in continuum and the emission or absorption coefficient in a spectral line, respectively:

\begin{equation}
\begin{array}{c}
\displaystyle
\varepsilon(z,\nu) = \varepsilon_c(\nu)+\frac{\displaystyle h\nu}{\displaystyle 4\pi}A_{ik} N n_i(z)\phi_{ik}(\nu), \nonumber\\ [10pt]
\displaystyle
\kappa(z,\nu) = \kappa_c(\nu) + \frac{\displaystyle \lambda^2}{\displaystyle 8\pi}A_{ik}N \left (\frac{\displaystyle g_i}{\displaystyle g_k}n_k(z)-n_i(z) \right) \phi_{ik}(\nu). \\[10pt]
\end{array}
\end{equation}

\noindent
Here, $\varepsilon_c(\nu)$ and $\kappa_c(\nu)$ are the dust emission and absorption coefficients in continuum, respectively; $A_{ik}$ is the Einstein coefficient for spontaneous emission; $n_i(z)$ and $n_k(z)$ are the normalized populations of levels $i$ and $k$, $\sum_{j} n_j(z) = 1$; $N$ is the particle number density; $g_i$ and $g_k$ are the statistical weights of the levels; $\lambda$ is the radiation wavelength; and $\phi_{ik}(\nu)$ is the normalized spectral line profile. In these formulas, it is implied that level $i$ lies above level $k$ in energy, i.e., $\varepsilon_i > \varepsilon_k$. The spectral profile of the emission and absorption coefficients $\phi_{ik}(\nu)$ is

\begin{equation}
\phi_{ik}(\nu)=\frac{1}{\sqrt{\pi}\Delta\nu_{ik}} \text{exp} \left(-\left(\frac{\nu-\nu_{ik}}{\Delta\nu_{ik}} \right)^2 \right),
\nonumber
\end{equation}

\noindent
where $\nu_{ik}$ is the transition frequency and $\Delta\nu_{ik}$ is the line profile width. The line profile width is determined by the spread in thermal velocities of the molecules and turbulent velocities in the gas-dust cloud:

\begin{equation}
\Delta\nu_{ik}=\nu_{ik}\frac{v_D}{c}, \quad v_D^{2}=v_T^2+v_{\text{turb}}^2,
\nonumber
\end{equation}

\noindent
where $v_T = \sqrt{2kT_g/m}$ is the most probable thermal velocity of the molecules, $k$ is the Boltzmann constant, $T_g$ is the gas kinetic temperature, $m$ is the mass of the molecule, and $v_{\text{turb}}$ is the characteristic turbulent velocity in the cloud.

\subsection*{The System of Statistical Equilibrium Equations for the Level Populations}

In the stationary case, the system of equations for the level populations is

\begin{equation}
\begin{array}{c}
\displaystyle
\sum_{k=1, \, k \ne i}^M \left( R_{ki}(z)+C_{ki} \right) n_k(z) - n_i(z)\sum_{k=1, \, k \ne i}^M \left( R_{ik}(z)+C_{ik} \right)=0, \quad i=1,...,M-1, \\
\displaystyle
\sum_{i=1}^M n_i(z)=1,
\end{array}
\label{eq5}
\end{equation}

\noindent
where $M$ is the total number of levels, $R_{ik}(z)$ is the rate coefficient for the transition from level $i$ to level $k$ through radiative processes, and $C_{ik}$ is the rate coefficient for the transition from level $i$ to level $k$ through collisional processes. The rate coefficients for radiative transitions $R_{ik}(z)$ are

\begin{equation}
\begin{array}{c}
R_{ik}^{\downarrow}(z)=B_{ik}J_{ik}(z)+A_{ik}, \quad \varepsilon_i > \varepsilon_k, \\[10pt]
R_{ik}^{\uparrow}(z)=B_{ik}J_{ik}(z), \quad \varepsilon_i < \varepsilon_k.
\end{array}
\nonumber
\end{equation}

\noindent
Here,$A_{ik}$ and $B_{ik}$ are the Einstein coefficients for spontaneous and stimulated emission, respectively; $J_{ik}(z)$ is the radiation intensity averaged over the direction and over the line profile:

\begin{equation}
J_{ik}(z)=\frac{1}{2} \int\limits_{-\infty}^{\infty} d\nu \, \phi_{ik}(\nu) \int\limits_{-1}^{1} d\mu  \,  I(z,\mu,\nu),
\nonumber
\end{equation}

\noindent
where $I(z,\mu,\nu)$ is the solution of Eq. (\ref{eq4}).

The collisional transitions of H$_2$O molecules in collisions with H and He atoms and H$_2$ molecules are considered in the model. If the level structure of the collisional partners is unimportant or if the level populations have the Boltzmann distribution, then the principle of detailed balance holds for the molecule's collisional excitation and de-excitation rate coefficients:

\begin{equation}
\displaystyle
C_{ik} = \frac{g_k}{g_i}C_{ki} \,\text{exp} \left( \frac{\varepsilon_i-\varepsilon_k}{kT_g} \right),
\nonumber
\end{equation}

\noindent
where the first index in the collisional rate coefficient denotes the initial level. In the case of a non-equilibrium level population distribution for the collisional partner, the rate coefficients for the collisional transitions of the H$_2$O molecule can be written as

\begin{equation}
C_{ik} = \sum_{l} \xi_{l} \sum_{m} C_{ik}^{lm}
\nonumber
\end{equation} 

\noindent
where $C_{ik}^{lm}$ is the rate coefficient for the transition of the H$_2$O molecule from level $i$ to level $k$ and the transition of the incident particle from level $l$ to level $m$ during the collision, $\xi_l$ are the normalized level populations of the incident particle. The state-to-state rate coefficients $C_{ik}^{lm}$ obey the principle of detailed balance:

\begin{equation}
\displaystyle
C_{ik}^{lm} = \frac{g_k}{g_i} \frac{g_m}{g_l} C_{ki}^{ml} \,\text{exp} \left( \frac{\varepsilon_i + \varepsilon_l - \varepsilon_k - \varepsilon_m}{kT_g} \right).
\nonumber
\end{equation}

\noindent
The sum $\displaystyle \sum_{m} C_{ik}^{lm}$ is called the effective collisional rate coefficient.

\subsection*{Collisional Rate Coefficients and Spectroscopic Data}
In our calculations, we took into account 45 ortho-H$_2$O rotational levels and 9 H$_2$ rotational levels. The spectroscopic data for H$_2$O were taken from the HITRAN 2012 database (Rothman et al. 2013). The H$_2$ energy levels were taken from Dabrowski (1984). 

We used the collisional rate coefficients for collisions between H$_2$O and H$_2$ molecules from Dubernet et al. (2009) and Daniel et al. (2010, 2011). In the calculations of these collisional rate coefficients, the interaction potential from Valiron et al. (2008) was used. The tables of rate coefficients are accessible in the BASECOL database\footnote{\url{http://basecol.obspm.fr/}} (Dubernet et al. 2006, 2013). Estimates of the effective collisional rate coefficients (summed over the final states of H$_2$) for H$_2$O de-excitation transitions for collisions with H$_2$ molecules in the states up to $J = 8$ are presented in the mentioned papers. We estimated the effective collisional rate coefficients for ortho-H$_2$O excitation (reverse) transitions in collisions with para-H$_2$ molecules based on data from Dubernet et al. (2009). The ortho-H$_2$ level populations in our calculations corresponded to the Boltzmann distribution. Therefore,
the total rate coefficients for ortho-H$_2$O excitation (reverse) transitions for collisions with ortho-H$_2$ molecules were calculated from the principle of detailed balance.

The collisional rate coefficients for transitions between H$_2$O levels in inelastic collisions of H$_2$O with He atoms were taken from Green et al. (1993). The rate coefficients for transitions between H$_2$O levels in collisions of H$_2$O with H atoms were assumed to be a factor of 1.2 larger than those for collisions of H$_2$O with He atoms. The numerical coefficient allows for the difference in H and He atom masses and cross sections.

\subsection*{The Dust Model}
In our calculations, we used the dust model from Weingartner and Draine (2001) and Draine (2003) \footnote{ \url{http://www.astro.princeton.edu/~draine/dust/dustmix.html}}. The dust emissivity was calculated in accordance with Kirchhoff's radiation law. At dust temperatures $T_d \lesssim 100$ K, the dust radiation has virtually no effect on the H$_2$O level populations (Yates et al. 1997). Here, we did not consider the thermal balance of dust and gas.

\subsection*{Numerical Calculations}
The radiative transfer equation in the medium (\ref{eq4}) and the statistical equilibrium equations for the molecular level populations (\ref{eq5}) are a system of non-linear equations for the level populations of the H$_2$O molecule. This system of equations was solved by the accelerated $\Lambda$-iteration method (Rybicki and Hummer 1991).

The cloud in our numerical model was broken down into layers parallel to the cloud plane. The molecular level populations within each layer were constant. The thickness of the near-surface layer was chosen in such a way that the optical depth for any H$_2$O line and any considered direction in the layer was less than 1. The thickness of each succeeding layer into the cloud was larger than that of the preceding one by a constant factor. The number of layers into which the cloud was broken down is 100. The number of unknown molecular level populations is $N \times M$, where $N$ is the number of cloud layers and $M$ is the number of molecular levels. The range of values for the parameter $\mu$ is $[0;1]$; the discretization step was chosen in our calculations to be 0.1. The deviation of the radiation frequency from the transition frequency is characterized by the parameter $x = (\nu - \nu_{ik})/\Delta \nu_{ik}$. The range of values for the parameter $x$ for each line was chosen to be $[-5;5]$; the discretization step is 0.25.

An additional acceleration of the iterative series was achieved by applying the convergence optimization method proposed by Ng (1974). The convergence criterion for the iterative series was the condition on the maximum relative increment in level populations for two successive iterations, $\displaystyle \max_i |\Delta n_i/n_i| < 10^{-4}$. In our calculations, we used the algorithms for solving systems of linear equations published in Rybicki and Hummer (1991) and in the book by Press et al. (1997). For a description of our calculations, see also Nesterenok (2013). 

The calculations were performed on the supercomputer of the St. Petersburg branch of the Joint Supercomputer Center, the Russian Academy of Sciences.

\section*{Results}
\subsection*{Ortho-H$_2$O Level Population Inversion in the 22.2-GHz Line}

The population inversion for levels $i$ and $j$ is

\begin{equation}
\Delta n_{ij}(z) = \frac{n_i(z)}{g_i} - \frac{n_j(z)}{g_j},
\nonumber
\end{equation}

\noindent
where $\varepsilon_i > \varepsilon_j$. The cloud-averaged population inversion for levels $i$ and $j$ is calculated from the formula

\begin{equation}
\langle \Delta n_{ij} \rangle = \frac{1}{H} \int \limits_0^H dz \, \Delta n_{ij}(z).
\nonumber
\end{equation} 

\noindent
The excitation temperature $T_\text{exc}^{ij}$ for two levels, $i$ and $j$, is determined from the equality

\begin{equation}
\frac{1}{T_\text{exc}^{ij}} = \frac{k}{\Delta \varepsilon_{ij}} \text{ln} \left( \frac{n_j g_i}{n_i g_j} \right),
\nonumber
\end{equation}   

\noindent
where $\Delta \varepsilon_{ij} = \varepsilon_i-\varepsilon_j$, with $i > j$. For the H$_2$ $J = 0$ and 2 levels, $\Delta \varepsilon_{20} = 354.35$ cm$^{-1}$. When the level excitation temperature $T_\text{exc}^{ij}$ is equal to the gas temperature $T_g$, the level populations correspond to the Boltzmann distribution for a given temperature. If the lower-level population $n_j \to 0$, then $1/T_\text{exc}^{ij} \to -\infty$; for $n_i/g_i = n_j/g_j$, $1/T_\text{exc}^{ij} = 0$; when the upper-level population $n_i \to 0$, $1/T_\text{exc}^{ij} \to +\infty$.

Figure 2 presents the results of our calculations of the ortho-H$_2$O $6_{16}$ and $5_{23}$ level population inversion (the 22.2-GHz line) as a function of distance into the cloud. The results of our calculations are presented for a gas temperature of 400 K. The H$_2$ level populations correspond to the Boltzmann distribution in this case. For distances $z/H \gtrsim 0.2$ into the cloud, the population inversion depends weakly on the coordinates. There is collisional maser pumping, with the sink line photons being absorbed by cold dust.

\begin{figure}[t]
\centering
\includegraphics[width = 0.7\textwidth]{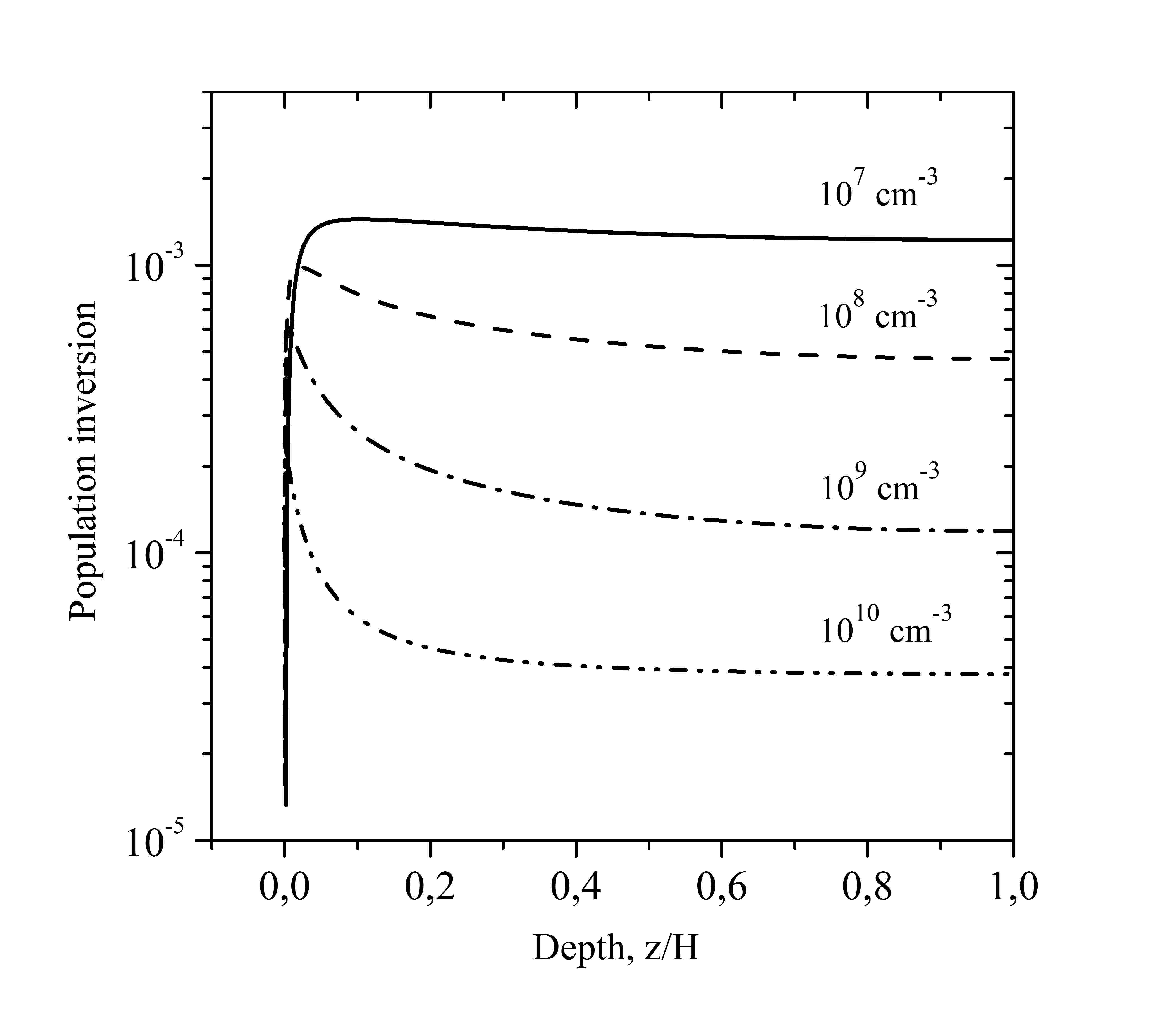}
\caption{\small{Ortho-H$_2$O $6_{16}$ and $5_{23}$ level population inversion (the 22.2-GHz line) versus distance into the cloud. The total number density of hydrogen atoms $N_\text{H,\,tot}$ is specified near each curve. The results of our calculations are presented for the gas temperature $T_g = 400$ K.}}
\end{figure}

The H$_2$ $J = 0$ and 2 level populations were varied in our calculations, with their sum having remained fixed and equal to the sum of the populations for the Boltzmann distribution. The populations of the remaining H$_2$ levels corresponded to the Boltzmann distribution. The H$_2$ level populations were assumed to be independent of the $z$ coordinate. For each distribution of H$_2$ level populations, i.e., for a fixed excitation temperature $T_\text{exc}^{20}$ for the $J = 0$ and 2 levels, we solved the system of non-linear equations (\ref{eq4}) and (\ref{eq5}) for the ortho-H$_2$O level populations. The H$_2$O level populations are functions of the parameter $T_\text{exc}^{20}$.

Figure 3 presents the results of our calculations of the cloud-averaged ortho-H$_2$O level population inversion $\langle \Delta n_{ij} \rangle$ for the 22.2-GHz transition. The results were normalized to the parameter $\langle \Delta n_{ij} \rangle$ for a thermodynamically equilibrium H$_2$ level population distribution ($T_\text{exc}^{20} = T_g$). For gas densities $N_\text{H,\,tot} \lesssim 10^8$ cm$^{-3}$, the $6_{16}$ and $5_{23}$ level population inversion increases with increasing H$_2$ $J = 2$ level population. Thus, the effect of quasi-resonance energy transfer in collisions between H$_2$O and H$_2$ leads to more efficient H$_2$O maser pumping. For high gas densities $N_\text{H,\,tot} \gtrsim 10^9$ cm$^{-3}$, this effect has the opposite sign. The effect of quasi-resonance energy transfer in collisions between H$_2$O and H$_2$ weakens with increasing gas temperature.

\begin{figure}[ht]
\centering
\includegraphics[width = 0.8\textwidth]{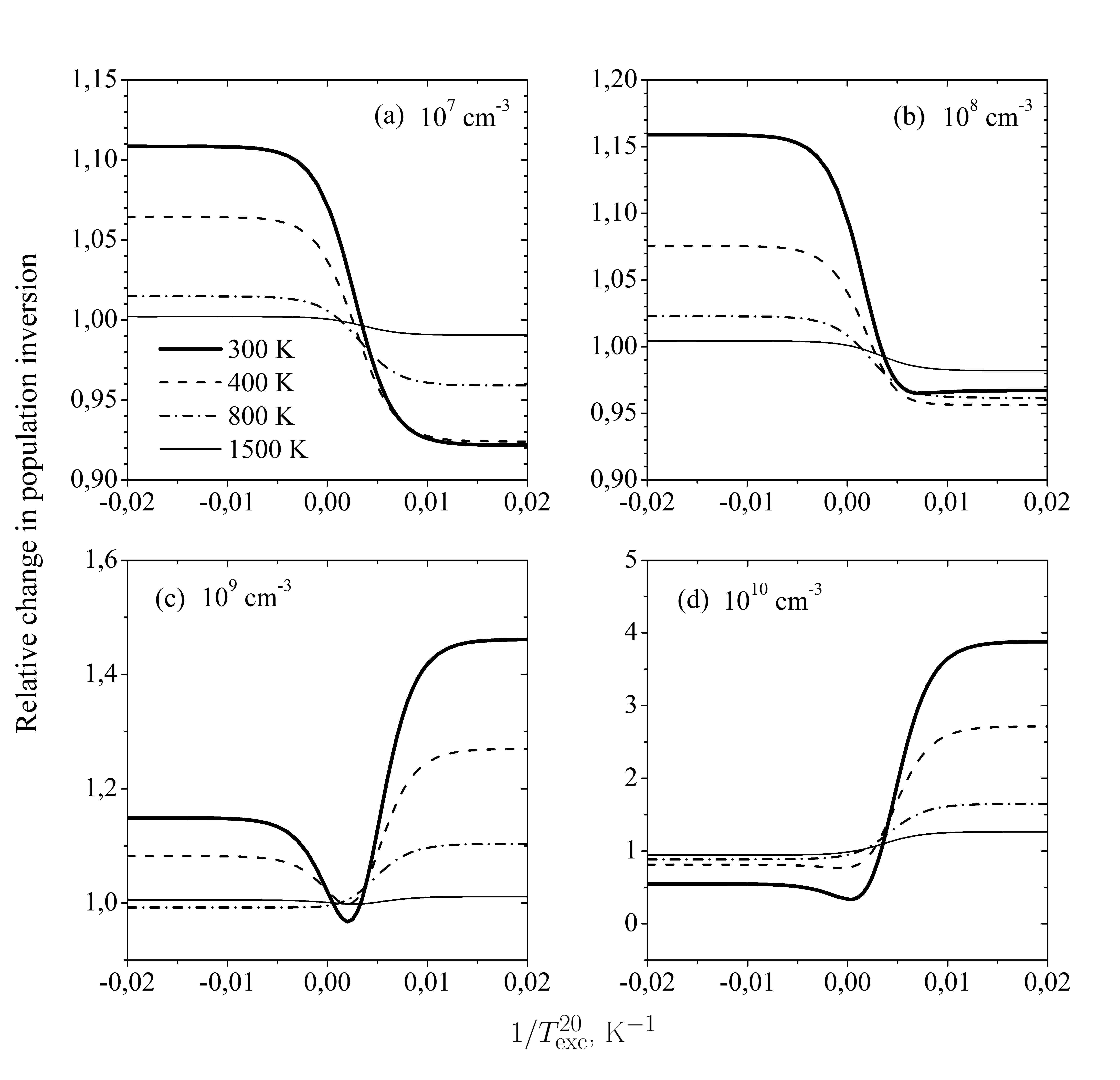}
\caption{\small{Relative change in ortho-H$_2$O $6_{16}$ and $5_{23}$ level population inversion (the 22.2-GHz line) $\langle \Delta n_{ij} \rangle / \langle \Delta n_{ij} (T_g)\rangle$ is along the vertical axis, where $\langle \Delta n_{ij} (T_g)\rangle$ is the value of the parameter in the case of a thermodynamically equilibrium H$_2$ level population distribution. The parameter $1/T_\text{exc}^{20}$ is along the horizontal axis, where $T_\text{exc}^{20}$ is the excitation temperature of the H$_2$ $J = 0$ and 2 levels. The total number density of hydrogen atoms $N_\text{H,\,tot}$ is specified on each plot.}}
\end{figure}

\subsection*{Ortho-H$_2$O Level Population Inversion in the 380-, 439-, and 621-GHz Lines}

The 380-, 439-, and 621-GHz lines correspond to the ortho-H$_2$O $4_{14} \to 3_{21}$, $6_{43} \to 5_{50}$, and $5_{32} \to 4_{41}$ transitions, respectively. The maser emission in the 439- and 621-GHz lines was observed in the envelopes of late-type stars and in star-forming regions in our Galaxy (Melnick et al. 1993; Neufeld et al. 2013). The 439-GHz line was also observed toward the galaxy NGC 3079 (Humphreys et al. 2005). 

Figure 4 presents the results of our calculations of the cloud-averaged H$_2$O level population inversion $\langle \Delta n_{ij} \rangle$ for the 380-, 439-, and 621-GHz transitions. The results of our calculations are presented for the gas number densities $N_\text{H,\,tot} = 10^8$ cm$^{-3}$, and $10^9$ cm$^{-3}$. For some values of the physical parameters, the level population inversion exists only at $1/T_\text{exc}^{20}$ below some critical value (i.e., for high H$_2$ $J = 2$ level population). Thus, there is quasi-resonance H$_2$O maser pumping. 

The H$_2$O level population inversion is proportional to the fraction of H$_2$ molecules at excited levels. For the gas temperature $T_g = 400$ K and under the condition $1/T^{20}_\text{exc} \to -\infty$, the H$_2$ $J = 2$ level population in our model is 0.2 (in this case, the $J = 0$ level population is 0). The level population inversion is also proportional to the number of quasi-resonance H$_2$O transitions involved in the maser pumping. The number of such transitions in our model is about 10-30 for the maser lines under consideration.

\begin{figure}[!ht]
\centering
\includegraphics[width = 0.85\textwidth]{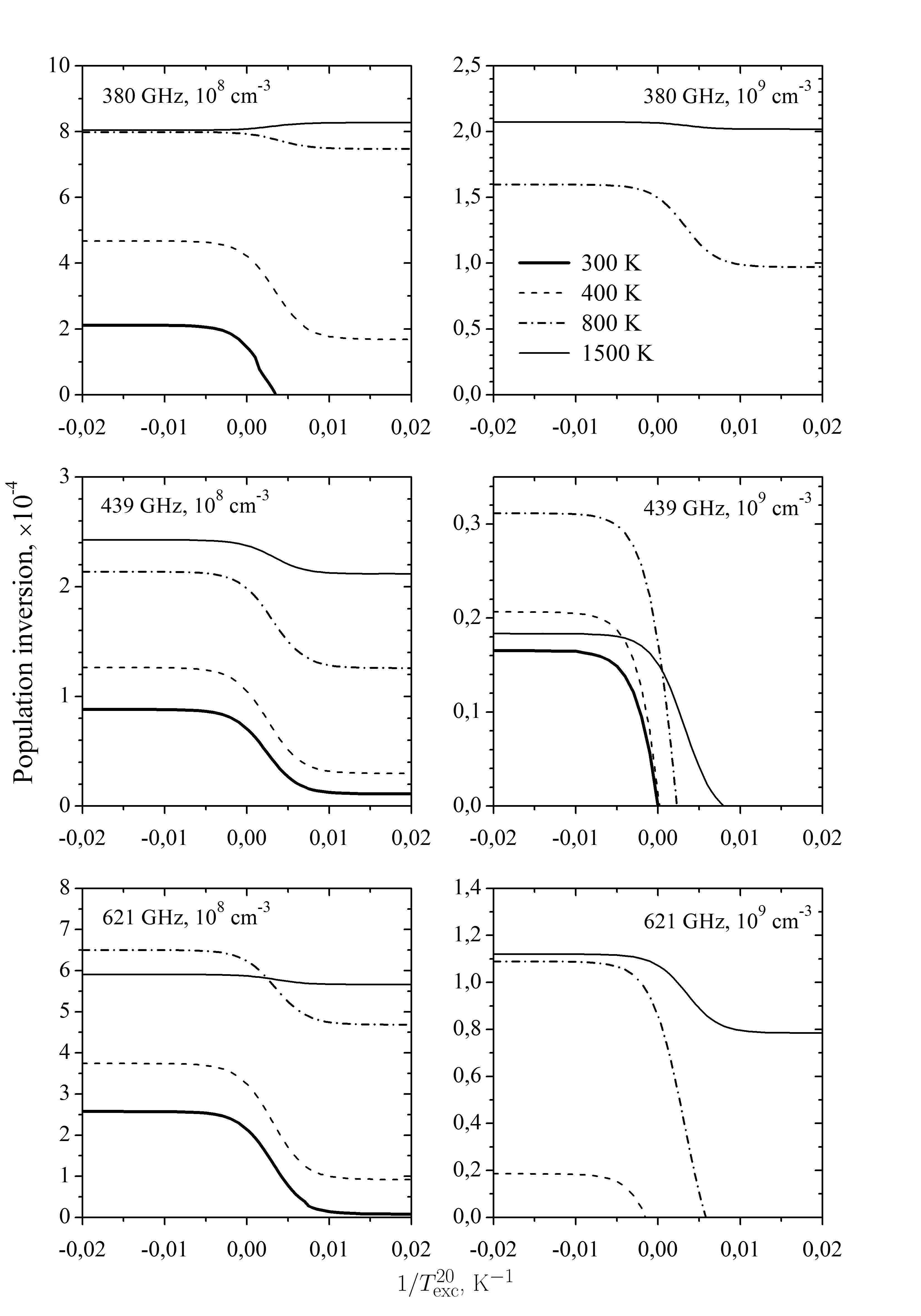}
\caption{\small{Ortho-H$_2$O $4_{14}$ and $3_{21}$ (the 380-GHz line), $6_{43}$ and $5_{50}$ (the 439-GHz line), $5_{32}$ and $4_{41}$ (the 621-GHz line) level population inversion versus parameter $1/T_\text{exc}^{20}$. The transition frequency and the total number density of hydrogen atoms $N_\text{H,\,tot}$ are specified on each plot.}}
\end{figure}

\section*{Discussion}

The turbulent motions of gas in molecular clouds, the expansion of compact HII regions, and the interaction of bipolar outflows from protostars and young stars with clouds of molecular gas give rise to shock waves. The possibility of the generation of intense H$_2$O maser emission in shocks in star-forming regions was considered in Strelnitskii (1973), Elitzur et al. (1989), Hollenbach et al. (2013), and other papers. In dissociative shocks, there is collisional dissociation of H$_2$ molecules in a hot gas at the shock front (Flower 2007). As the distance from the shock front increases, the gas temperature decreases and H$_2$ molecules are formed on dust particles. The gas temperature in this region of the molecular flow is 300-400 K and the generation of H$_2$O maser emission is possible (Hollenbach et al. 2013). The H$_2$ molecules formed on dust particles are at excited rotational-vibrational levels and there can be quasi-resonance energy transfer in collisions between H$_2$ and H$_2$O molecules.

The central engines of AGNs are powerful sources of X-ray emission. Because of the high photon energy and the small absorption cross section, the X-ray emission exerts a deeply penetrating action on the physical properties of the interstellar medium (Malony et al. 1996). To all appearances, the H$_2$O maser emission in the central regions of AGNs is generated in clouds of gas and dust, in which the physical conditions are governed by the X-ray emission from the central engine, i.e., in X-ray dissociation regions (Neufeld et al. 1994; Collison and Watson 1995; Malony 2002). In X-ray dissociation regions, the interaction of high-energy photoelectrons with atoms and molecules of the gas leads to the excitation of electronic and vibrational states of molecules. In addition, the H$_2$ molecules formed on dust particles are at highly excited rotational-vibrational levels.

One of the main mechanisms for the excitation of high-energy rotational-vibrational H$_2$ levels is the formation of molecules on dust particles. For gas number densities $N_\text{H,\,tot} \gtrsim 10^5$ cm$^{-3}$, the de-excitation of excited H$_2$ levels occurs through collisions with H atoms, provided that $N_\text{H} \sim N_{\text{H}_2}$ (Hollenbach and Tielens 1999). Let $n_\text{x}$ be the fraction of H$_2$ molecules at highly excited rotational-vibrational levels. In the stationary case, the excitation rate of high-energy molecular levels is equal to the de-excitation rate. We can then write

\begin{equation}
R_\text{gr} N_\text{H,\,tot} N_\text{H} = C_\text{x} n_\text{x} N_{\text{H}_2} N_\text{H},
\label{eq6} 
\end{equation}

\noindent
where $R_\text{gr}$ is the specific formation rate of H$_2$ molecules on dust and $C_\text{x}$ is the mean collisional de-excitation rate of excited H$_2$ levels. For our estimates, we take $R_\text{gr}$ to be $3\times 10^{-17}$ cm$^3$ s$^{-1}$ (Bourlot et al. 2012). Assuming that the H$_2$ molecule is de-excited after 3-5 collisions and that $T_g = 400$ K, we have an estimate of $\sim 10^{-11}$ cm$^3$ s$^{-1}$ for $C_\text{x}$ (Wrathmall et al. 2007). From Eq. (\ref{eq6}) we have

\begin{equation}
n_\text{x} \simeq 10^{-5} \left(\frac{0.3}{x_{\text{H}_2}} \right),
\nonumber
\end{equation}  

\noindent
where $x_{\text{H}_2} = N_{\text{H}_2}/N_\text{H,\,tot}$. The number of H$_2$O and H$_2$ transitions for which the quasi-resonance conditions are met and which can be involved in the H$_2$O maser pumping is 10$^4$ -- 10$^5$. When estimating this number, we took into account 318 H$_2$ rotational-vibrational levels and the possibility of H$_2$O vibrational state excitation. A large number of quasi-resonance transitions can lead to a noticeable effect, despite the relatively small fraction of H$_2$ molecules in highly excited states.

\section*{Conclusions}

We considered the effect of quasi-resonance energy transfer in collisions between H$_2$ and H$_2$O molecules on the H$_2$O maser pumping process. We calculated the populations of 45 ortho-H$_2$O rotational levels for various populations of the para-H$_2$ $J = 0$ and 2 rotational levels. The H$_2$O level population inversion in the 22.2-, 380-, 439-, and 621-GHz maser lines was shown to depend significantly on the population distribution of the lower H$_2$ rotational levels. For some of the physical parameters, the effect of quasi-resonance energy transfer between molecules leads to a high H$_2$O level population inversion in lines of the submillimeter wavelength range; while the H$_2$O level population inversion in the model with a thermodynamically equilibrium H$_2$ level population distribution is either small or absent. The quasi-resonance energy transfer effect can play a significant role in pumping H$_2$O masers in the central regions of AGNs and in star-forming regions.

\section*{Acknowledgements} 

This work was supported in part by the Russian Foundation for Basic Research (project no. 14-02-31302), the Program of the President of Russia for Support of Leading Scientific Schools (project no. NSh-294.2014.2), and the Research Program OFN-17, the Division of Physics of the Russian Academy of Sciences.

\section*{References}

\noindent
1. Babkovskaia N. and Poutanen J., Astron. Astrophys. {\bf 418}, 117 (2004).

\noindent
2. Bolgova G.T., Strelnitskii V.S., Shmeld I.K., Soviet Astronomy {\bf 21}, 468 (1977).

\noindent
3. Castangia P., Impellizzeri C.M.V., McKean J.P., Henkel C., Brunthaler A., Roy A.L., Wucknitz O., Ott J., et al., Astron. Astrophys. {\bf 529}, A150 (2011).

\noindent
4. Churchwell E., Witzel A., Huchtmeier W., Pauliny-Toth I., Roland J., and Sieber W., Astron. Astrophys. {\bf 54}, 969 (1977).

\noindent
5. Collison A.J. and Watson W.D., Astrophys. J. {\bf 452}, L103 (1995).

\noindent
6. Dabrowski I., Canad. J. Phys. {\bf 62}, 1639 (1984).

\noindent
7. Daniel F., Dubernet M.-L., Pacaud  F., and Grosjean  A., Astron. Astrophys. {\bf 517}, A13 (2010).

\noindent
8. Daniel F., Dubernet M.-L., and Grosjean  A., Astron. Astrophys. {\bf 536}, A76 (2011).

\noindent
9. dos Santos P.M. and L$\acute{e}$pine J.R.D., Nature {\bf 278}, 34 (1979).

\noindent
10. Draine B.T., Ann. Rev. Astron. Astrophys. {\bf 41}, 241 (2003).

\noindent
11. Dubernet M.-L., Grosjean A., Flower D., Roueff E., Daniel F., Moreau N., and Debray B., J. Plasma Fusion Res. Series {\bf 7}, 356 (2006).

\noindent
12. Dubernet M.-L., Daniel F., Grosjean A., and Lin C.Y., Astron. Astrophys. {\bf 497}, 911 (2009).

\noindent
13. Dubernet M.-L., Alexander M.H., Ba Y.A., Balakrishnan N., Balan\c{c}a C., Ceccarelli C., Cernicharo J., Daniel F., et al., Astron. Astrophys. {\bf 553}, A50 (2013).

\noindent
14. Elitzur M., Hollenbach D.J., and McKee C.F., Astrophys. J. {\bf 346}, 983 (1989).

\noindent
15. Flower D.R., Molecular collisions in the interstellar medium (New York: Cambridge University Press, 2007).

\noindent
16. Green S., Maluendes S., and McLean A.D., Astrophys. J. Suppl. Ser. {\bf 85}, 181 (1993).

\noindent
17. Greenhill L.J., Booth R.S., Ellingsen S.P., Herrnstein J.R., Jauncey D.L., McCulloch P.M., Moran J.M., Norris R.P., et al., Astrophys. J. {\bf 590}, 162 (2003).

\noindent
18. Hollenbach D.J. and Tielens A.G.G.M., Reviews of Modern Physics {\bf 71}, 173 (1999).

\noindent
19. Hollenbach D., Elitzur M., and McKee C.F., Astrophys. J. {\bf 773}, 70 (2013).

\noindent
20. Humphreys E.M.L., Greenhill L.J., Reid M.J., Beuther H., Moran J.M., Gurwell M., Wilner D.J., and Kondratko P.T., Astrophys. J. {\bf 634}, L133 (2005). 

\noindent
21. Kartje J.F., K$\ddot{o}$nigl A., and Elitzur M., Astrophys. J. {\bf 513}, 180 (1999).

\noindent
22. Le Bourlot J., Le Petit F., Pinto C., Roueff E., and Roy F., Astron. Astrophys. {\bf 541}, A76 (2012).

\noindent
23. Liljestr$\ddot{o}$m T., Mattila K., Toriseva M., and Anttila R., Astron. Astrophys. Suppl. Ser. {\bf 79}, 19 (1989).

\noindent
24. Maloney P.R., Hollenbach D.J., and Tielens A.G.G.M., Astrophys. J. {\bf 466}, 561 (1996).

\noindent
25. Maloney P.R., Publ. Astron. Soc. Aust. {\bf19}, 401 (2002).

\noindent
26. Matveenko L.I., Diamond P.J., Graham D.A., Astronomy Reports {\bf 44}, 592 (2000).

\noindent
27. Matveenko L.I., Demichev V.A., Astronomy Reports {\bf 54}, 986 (2010).

\noindent
28. Melnick G.J., Menten K.M., Phillips T.G., and Hunter T., Astrophys. J. {\bf 416}, L37 (1993).

\noindent
29. Moran J.M., Frontiers of Astrophysics: A Celebration of NRAO's 50th Anniversary, ASP Conference Series 395 (Ed. Bridle A.H., Condon J.J., and Hunt G.C., Charlottesville, 2008), p.87.

\noindent
30. Nesterenok A.V., Varshalovich D.A., Astronomy Letters {\bf 37}, 456 (2011).

\noindent
31. Nesterenok A.V., Astronomy Letters {\bf 39}, 717 (2013).

\noindent
32. Neufeld D.A., Maloney P.R., and Conger S., Astrophys. J. {\bf 436}, L127 (1994).

\noindent
33. Neufeld D.A., Wu Y., Kraus A., Menten K.M., Tolls V., Melnick G.J., and Nagy Z., Astrophys. J. {\bf 769}, 48 (2013). 

\noindent
34. Ng K.-C., J. Chem. Phys. {\bf 61}, 2680 (1974).

\noindent
35. Palagi F., Cesaroni R., Comoretto G., Felli M., and Natale V., Astron. Astrophys. Suppl. Ser. {\bf 101}, 153 (1993).

\noindent
36. Peck A.B., Henkel C., Ulvestad J.S., Brunthaler A., Falcke H., Elitzur M., Menten K.M., and Gallimore J.F., Astrophys. J. {\bf 590}, 149 (2003). 

\noindent
37. Press W.H., Teukolsky S.A., Vetterling W.T., and Flannery B.P., Numerical Recipes in C. The Art of Scientific Computing (Cambridge: Cambridge Univ. Press, 1997).

\noindent
38. Rothman L.S.,Gordon I.E., Babikov Y., Barbe A., Chris Benner D., Bernath P.F., Birk M., Bizzocchi L., et al., J. Quant. Spectrosc. Rad. Transfer {\bf 130}, 4 (2013).

\noindent
39. Rybicki G.B. and Hummer D.G., Astron. Astrophys. {\bf 245}, 171 (1991).

\noindent
40. Strelnitskii V.S., Sov. Astron. {\bf 17}, 717 (1973).

\noindent
41. Tarchi A., Cosmic Masers - from OH to H$_0$, Proceedings IAU Symposium {\bf 287}, 323 (2012).

\noindent
42. Tennyson J., Zobov N.F., Williamson R., and Polyansky O.L., J. of Physical and Chemical Reference Data {\bf 30}, 735 (2001).

\noindent
43. Torrelles J.M., Patel N., G$\acute{o}$mez J.F., Anglada G., and Uscanga L., Astrophys. Space Sci. {\bf 295}, 53 (2005).

\noindent
44. Uscanga L., Cant$\acute{o}$ J., Curiel S., Anglada G., Torrelles J.M., Patel N.A., G$\acute{o}$mez J.F., and Raga A.C., Astrophys. J. {\bf 634}, 468 (2005).

\noindent
45. Valiron P., Wernli M., Faure A., Wiesenfeld L., Rist C., Ked$\check{z}$uch S., and Noga J., J. Chem. Phys. {\bf 129}, 134306 (2008).

\noindent
46. Varshalovich D.A., Kegel W.K., Chandra S., Soviet Astronomy Letters {\bf 9}, 209 (1983).

\noindent
47. Weingartner J.C. and Draine B.T., Astrophys. J. {\bf 548}, 296 (2001).

\noindent
48. Wrathmall S.A., Gusdorf  A., and Flower D.R., Mon. Not. R. Astron. Soc. {\bf 382}, 133 (2007).

\noindent
49. Yates J.A., Field D., and Gray M.D., Mon. Not. R. Astron. Soc. {\bf 285}, 303 (1997).

\end{document}